\documentclass{amsproc}
\usepackage{color}
\usepackage{graphics}
\usepackage{graphicx}
\usepackage{epsfig}
\usepackage{amssymb}
\usepackage{verbatim}
\usepackage{graphics,amsmath,amsfonts,amssymb}
\usepackage{epsfig,verbatim}
\usepackage{cite}
\usepackage{lineno,hyperref}
\usepackage{amsmath}%
\usepackage{amsfonts}%
\usepackage{amssymb}%
\usepackage{graphicx}
\usepackage[all]{xy}
\newtheorem{theorem}{Theorem}[section]

\theoremstyle{definition}

\theoremstyle{remark}
\newtheorem{remark}[theorem]{Remark}

\numberwithin{equation}{section}

\newcommand{\vx}{\left({\bf x}\right)}
\newcommand{\argT}{\left( t \right)}
\newcommand{\lp}{\left(}
\newcommand{\rp}{\right)}

\begin{document}
\title[D. Galois Theory, Quivers and Seismic]{Differential Galois Groups and Representations of Quivers for Seismic Models with Constant Hessian of Square of Slowness$^\dagger$\footnote{$\dagger$ Supported by COLCIENCIAS-ECOPETROL project (Contract RC. No. 0266-2013) and CODI (Estrategia de Sostenibilidad 2016-2017, Universidad de Antioquia UdeA)}}

\author[P. Acosta-Hum\'anez]{Primitivo Acosta-Hum\'anez}
\address[P. Acosta-Hum\'anez]{School of Basic and Biomedical Sciences,  \href{www.unisimonbolivar.edu.co}{Universidad Sim\'on Bol\'{\i}var}\\  Barranquilla, Colombia.}
\email{primitivo.acosta@unisimonbolivar.edu.co--primi@intelectual.co}

\author[H. Giraldo]{Hern\'an Giraldo}
\address[H. Giraldo]{Institute of Mathematics, \href{www.udea.edu.co}{Universidad de Antioquia}\\  Medell\'{\i}n, Colombia.}
\email{hernan.giraldo@udea.edu.co}

\author[C. Piedrahita]{Carlos Piedrahita$^*$\footnote{$*$ Corresponding author}}
\address[C. Piedrahita]{Department of Basic Sciences, \href{www.udem.edu.co}{Universidad de Medell\'{\i}n}\\
	Medell\'{\i}n, Colombia.}
\email{cpiedrahita@udem.edu.co}

\maketitle

\begin{abstract}
The trajectory of energy is modeled by the solution of the Eikonal equation, which can be solved by solving a  Hamiltonian system. This system is amenable of treatment from the point view of the theory of Differential Algebra. In particular, by Morales-Ramis theory it is possible to analyze integrable Hamiltonian systems through the abelian structure of their variational equations. In this paper we obtain the abelian differential Galois groups and the representation of the quiver, that allow us to obtain such abelian differential Galois groups, for some seismic models with constant Hessian of square of slowness, proposed in \cite{cerveny2001}, which are equivalent to linear Hamiltonian systems with three uncoupled harmonic oscillators. \newline

\noindent{\footnotesize {\textbf{Keywords and Phrases}. Differential Galois Theory, Eikonal Equation,  Hamilton Equation, Helmholtz Equation, High Frequency Approximation, Morales-Ramis Theory, Ray Theory, Representations of Quivers}\newline
}

\end{abstract}

\section{Introduction}
Wave equations appear in many areas of sciences, in particular, in seismology and seismic exploration. Asymptotic solutions give us an estimation to the actual solution to cases where it is difficult to obtain exact solutions. In this context, the idea of Ray theory is a very important tool used in geophysics. High frequency approximations or ray theory have been used in seismology to obtain good approximation for the solution of the acoustic and/or elastic wave partial differential equations (PDE). They represent the transmission of energy in a continuous media. The acoustic wave equation and Transport equation are second order linear PDE. The first one is transformed into the \emph{Eikonal equation}, which is a first order nonlinear PDE. The trajectory of energy is modeled by the solution of the Eikonal Equation and it can be solved by solving a Hamiltonian system. For more details, see \cite{bleistein1986}.\\

\noindent Galois theory in the context of linear differential equations is known as \emph{differential Galois theory} or also as  \emph{Picard-Vessiot  theory}, see \cite{ka,mag,marram,vasi,si2}. There are effective algorithms and procedures derived from differential Galois theory such as Kovacic Algorithm (see \cite{ko}), Kimura's Table (see \cite{ki}) and \emph{Algebrization Procedure} (see \cite{acthesis,acbook,acsiam,acmowe}) that allow us to obtain explicit solutions, whenever it can be possible, of  differential equations. Differential Galois theory has been applied successfully in mathematical physics. One of these applications corresponds to Morales-Ramis theory (see \cite{mo}), that is, the differential Galois theory linked with the non-integrability of dynamical systems, being the starting point to prove non-integrability of Hamiltonian and non-Hamiltonian systems, see \cite{acsiam,aabd,acalde,acbl2,acbl,acblva}. Similarly, differential Galois theory has been used to study integrability and non-integrability of polynomial vector fields (see \cite{almp,acpan}), integrability in quantum mechanics  (see \cite{acthesis,acbook,acmowe,acpan,acsu,acsu2}) and integrability in quantum optics (see \cite{akss}).\\

\noindent In the sixties and seventies P. Gabriel, in \cite{Ga}, showed a correspondence between algebras of finite dimension over algebraically closed fields with quivers. After that, started a great development of the theory of representation of finite dimensional algebras and representation of quivers. References for this are \cite{ARS}, \cite{ASS2006}, \cite{Ga1}, \cite{Gus}, \cite{Re1} and \cite{Rin1984}. \\

\noindent The connection between the representations of quivers (directed graphs)  and differential Galois group is that both try to find the solutions to systems of differential equations. Representation of quivers is used with great success in the theory of representations of finite dimensional algebras. In particular, using representations we can solve a system of $m$ differential equations of degree $s$ in $n$ indeterminates (see \cite{Pen1998}, Section 1). Moreover, we can see the representation theory of algebras as a generalization of linear algebra, in the sense that in linear algebra we have a single linear transformation acting on a vector space, while in representation theory we have simultaneously a finite number of linear transformations acting on a vector space.\\

\noindent In this paper we study some seismic problems with a  constant Hessian of square of slowness without interactions between positions by using Morales-Ramis theory and representation theory of quivers. Firstly, we write the equivalent Hamiltonian to the Eikonal equation with constant Hessian of the square of slowness without interactions between the positions, which is equivalent to study a linear Hamiltonian system of three uncoupled oscillators. After, due to the linearisation of the Hamiltonian vector field, the so-called variational equation, is equivalent to the linear Hamiltonian system. Then we compute the differential Galois groups and the representation of the quiver, that allow us to obtain such abelian differential Galois groups, for such systems.

\section{Preliminaries}
In this section, which is divided in three parts, we give the basic theoretical background to understand this article. In the first part we present a brief review of ray theory applied to seismic exploration. The second part contains a brief summary of differential Galois theory, which is Galois theory in the context of linear differential equations, including the Morales-Ramis theory to study the integrability of Hamiltonian Systems. The third part is devoted to an elementary background about theory of representations of quivers, which will be related with these Ray equations and differential Galois theory to explore new relations and obtain results that have  a mathematical interest.

\subsection{Ray theory}

Ray theory is a collection of theorems that have been developed in the field of asymptotic analysis. It is  very useful as an approximation to the solution of ODEs or PDEs, in the context of the so called high-frequency approximation. Essentialy, we suppose an asymptotic expansion of the unknown solution, see \cite{bleistein1986,Evans2010,Zwors2012}, in the frecuency domain as

\begin{eqnarray}\label{highFrequency}
{u}\left({\bf x},\omega \right) &\sim& \omega^{\beta}\exp^{i\omega\tau\left({\bf
		x}\right)}\sum_{j=0}^{\infty}\frac{A_j\left({\bf x}\right)}{\left(i\omega\right)^j} \\
\tau\left({\bf x}\right) &\equiv& \mbox{Transport time}, \nonumber \\
A_j\left({\bf x}\right) &\equiv& \mbox{Independent width of frequency}, \nonumber \\
&& j=0,1,2,\ldots \nonumber
\end{eqnarray}

The physical interpretation of the series \eqref{highFrequency} will be analyzed: the inverse Fourier transform

\begin{eqnarray}\label{solutionTransform}
\omega^{\beta}\exp^{i\omega\tau\left({\bf x}\right)} \rightarrow F\lp t - \tau\vx \rp,
\end{eqnarray}
is the frequency transform of a progressive wave, in which $F\lp t\rp$, is the inverse transform of $ \omega^{\beta} $, which propagates in the direction of increment of $ \tau \vx $.\\

\noindent In the series \eqref{highFrequency}, the terms that are divided by higher powers of $ \lp i\omega \rp $ represents successive integrations in the time domain, which can be interpreted as the smoothest terms of the original function, $F$. In the time domain, we can interpret this as an infinite chain of distributions each more smooth that the one before, added together. For example, consider the chain of distributions

\begin{eqnarray}\label{smoothings} \nonumber
\delta\lp t \rp \rightarrow H \argT \rightarrow r \argT,
\end{eqnarray}

where $\delta\lp t \rp$, represents the delta function, $H\argT$, represents the Heaviside distribution and 
$r \argT$, represents the ramp function. The Fourier transform of the first term represents the portion more singular of the solution, associated with the high frequencies, while the subsequent terms, represent the smoothest contributions, i.e., the low frequency terms of the solution of the Helmholtz's Equation 

\begin{eqnarray}\label{operatorForm}
LP = \left[\nabla^2 + \frac{\omega^2}{c^2\vx}\right]P\lp{\bf x},\omega\rp = 0.
\end{eqnarray}

Considering the wave equation in the frequency domain, i.e., the Helmholtz Equation, we substitute the asymptotic series. A formal definition of an asymptotic expansion can be found in \cite{bleistein1984}. Substituting this series in the Helmholtz Equation we obtain:

\begin{eqnarray}\label{asymptoticSubstitution}
LP &=& \omega^{\beta}\exp{i\omega\tau}\sum_{j=0}^{\infty}\frac{1}{{\lp i\omega \rp}^j}
\left[\omega^2 \left\{ \frac{1}{c^2}- {\lp \nabla \rp}^2 \right\}A_j \right.  \\
& & \left. + i\omega \left\{ 2\nabla\tau \cdot \nabla{A_j} + A_j{\nabla}^2\tau \right\} + {\nabla}^2 A_j \right].
\nonumber
\end{eqnarray}
The equation \eqref{asymptoticSubstitution} is satisfied whether the coefficients of the series to be equal to zero, see  \cite{cerveny2001}, therefore equating to zero each power of $i \omega$,  starting with the maximum power, $\beta + 2$, when $j=0$, we obtain the so called {\it Eikonal Equation}

\begin{eqnarray}\label{eikonal}
\lp \nabla\tau\vx \rp^2 -\frac{1}{c^2\vx} = 0.
\end{eqnarray}
For the series \eqref{asymptoticSubstitution}, the highest non-zero power of $\omega$ is $\beta + 1$, which is obtained when
$j=0$, and thus we obtain the so called Transport Equation of order zero 

\begin{eqnarray}\label{transportOrderZero}
2\nabla\tau\vx \cdot \nabla A_0\vx + A_0\nabla^2\tau\vx = 0.
\end{eqnarray}

The Eikonal Equation represents the travel time and geometrically, represents the wave fronts of the wave. If we consider the Eikonal Equation, we can solve it by the method of characteristics, see \cite{john1982}. In this way, the first order non-linear partial differential equation is converted to a system of ordinary differential equations, which corresponds to a Hamiltonian system, see \cite{arnold1989}.

Consequently, we obtain trajectories defined by the following system, see \cite{bleistein1986,cerveny2001,Rauch2012}:

\begin{eqnarray}\label{raySystemEquationsLambdaHalf}
\frac{d \textbf{x}}{d \sigma} &=& \textbf{p} \\ \nonumber
\\ \nonumber
\frac{d\textbf{p}}{d\sigma}&=&\frac{1}{2}\nabla\left(\frac{1}{c^{2}(\textbf{x})}\right)
= \frac{-\nabla c(\textbf{x})}{c^{3}(\textbf{x})} \\ \nonumber
\\ \nonumber
\frac{d\tau}{d\sigma}&=& \frac{1}{c^{2}(\textbf{x})}. \\ \nonumber
\end{eqnarray}

These trajectories represent the flow of energy The physical interpretation of the variables of the system \eqref{raySystemEquationsLambdaHalf} is the following, see \cite{bleistein1984}: $x_i$, represents the coordinates along the ray and  the $p_i$ represents the coordinate of the so called slowness vector. This vector has units of the inverse of the velocity of the propagation of the wave in the media. The variable $\tau$ represents the travel time along the ray and $\sigma$ represents an integration parameter, which can be selected in different ways according to its convenience. The physical ray is the vector $x_i, i=1,2,3$. Finally, the function $c(\textbf{x})$ represents the velocity field. It is an smooth function that represents the variations of the wave propagation along a medium, and it is an scalar function defined in a domain of $\mathbb{R}^3$, which represents macroscopically the structure of the medium. The scalar field $c^{-1}(\textbf{x})$ is called the \emph{slowness} and $c^{-2}(\textbf{x})$ is called the \emph{square of slowness},  see \cite{cerveny2001}.

The ray is therefore a path in space, where energy is propagated between a source and a receiver, and it is a good approximation when the length of  the wave is small compared to the object of interest. It is usually called the \emph{high frequency approximation}, see \cite{schleicher2007}.

The system just described, is the one to which we will apply the concepts of differential algebra to analyze the integrability of the ODE system that appears in seismology.

In particular, considering the 2D subcase corresponding to the plane $x_1x_3$, that is, setting $x_2=0$ and for instance $\textbf{x}=(x_1,0,x_3)$, the acoustic wave equation is given by   $$\nabla^2P=\frac{1}{c^2}\frac{\partial^2P}{\partial t^2},$$ which can be written as  $$\frac{\partial^2 P}{\partial x_1^2}+\frac{\partial^2P}{\partial x_3^2}=\frac{1}{c^2(\textbf{x},t)}\frac{\partial ^2 P}{\partial t^2},$$ being 
$P(\textbf{x},t)$ the pressure in the high frequency approximation. Thus, 
$$P(\textbf{x},t)=A(\textbf{x},t)e^{-i\tau (\textbf{x},t)},\,\, w\gg 1,\,\,(equivalently \,\, \lambda\ll 1).$$
Therefore the Eikonal equation \eqref{eikonal} in this case can be written as
\begin{equation}\label{eikonal2}
\left(\frac{\partial\tau}{\partial q_1}\right)^2+\left(\frac{\partial\tau}{\partial q_3}\right)^2=\frac{1}{c^2(\textbf{q})},\quad p^2_1+p^2_3-\frac{1}{c^2}=0,\end{equation} where $\textbf{q}=(q_1,q_2,q_3)$, $q_i=x_i$,  $p_i=\frac{\partial \tau}{\partial q_i}$ and $c(\textbf{q})$ is the velocity field, being $q_2=p_2=0$. In the context of Hamiltonian mechanics the vector $\textbf{q}$ represents the generalized coordinates and the vector $\textbf{p}$ represents the generalized momentum, see \cite{arnold1989}. We recall that the Hamiltonian corresponding to the Eikonal equation \eqref{eikonal2} has two degrees of freedom and it is a particular case of a Hamiltonian with three degrees of freedom ($q_2,\,p_2\in\mathbb{C}$). Moreover, we will study the integrability of quadratic Hamiltonian with three degrees of freedom without interaction of the positions, which are equivalent to seismic models with constant Hessian of square of slowness.

Finally, in a quadratic Hamiltonian with three degrees of freedom the only possible interactions of the positions are given by $q_1q_2$, $q_2q_1$, $q_1q_3$, $q_3q_1$, $q_2q_3$ and $q_3q_2.$ Although the Hessian of quadratic hamiltonians with interaction in the positions is constant, we will not consider this case in this paper.

\subsection{Differential Galois and Morales-Ramis Theories}

Picard-Vessiot theory is the Galois theory of linear differential
equations, which is also known as differential Galois theory. Although we will review here some of its main definitions and results, we refer the reader to~\cite{vasi} for a wide theoretical background. See also  \cite{acthesis,acbook} for short summaries of this theory. \\

\noindent By $M_n(k)$ we means the set of $n\times n$ square matrices with entries in the field $k$. We recall that the general linear group over $\mathbb{C}$ is given by $$\mathrm{GL}_n(\mathbb{C})=\{A\in M_n({\mathbb{C}}):\,\, \det(A)\neq 0\}.$$ Therefore, an algebraic group of matrices $2\times 2$ is a subgroup $G\subset
\mathrm{GL}_2(\mathbb{C})\subset M_{2}(\mathbb{C})$ defined by means of algebraic equations
in its matrix elements and in the inverse of its determinant. That
is, there exists a set of polynomials $P_i \in \mathbb C[x_1,\ldots,x_5]$,
for $i\in I\subseteq \mathbb{N}$, such that $A\in\mathrm{GL}_2(\mathbb{C})$ given by
\[
A=\left(\begin{array}{cc} x_{1} & x_{2} \\ x_{3} & x_{4}
\end{array}\right),
\]
belongs to $G$ if and only if  $P_i\left( x_{1},x_{2},x_{3},x_{4},x_5\right) = 0$ for all $i\in I\subseteq \mathbb{N}$ and where $x_5=\left(\det A \right)^{-1}$. It is said that $G$ is an algebraic manifold endowed with a
group structure. Recall that $A\in {\rm SL}_2(\mathbb{C})$ iff $A \in {\rm GL}_2(\mathbb{C})$ and $\det A=1$.\\

\noindent A group $G$ is called \emph{solvable} if and only if there
exists a chain of normal subgroups
\[
e=G_0\triangleleft G_1 \triangleleft \ldots \triangleleft G_n=G,
\]
satisfying that the quotient $G_i/G_j$ is abelian for all $n\geq i\geq j\geq 0$.\\

\noindent It is well known that any algebraic group $G$ has a unique connected normal
algebraic subgroup $G^0$ of finite index. In particular, the
\emph{identity connected component $G^0$} of $G$ is defined as the largest connected
algebraic subgroup of $G$ containing the identity. In case that $G=G^0$ we say that $G$ is a
\emph{connected group}. Moreover, if $G^0$ is solvable we say that $G$ is \emph{virtually solvable}.
\\

\noindent The following result provides the relationship between virtual solvability of an algebraic group
and its structure.

\begin{theorem}[Lie-Kolchin]\label{LiKo}
	Let $G\subseteq \mathrm{GL}_2(\mathbb{C})$ be a virtually solvable group.
	Then, $G^0$ is triangularizable, i.e., it is conjugate to a subgroup of upper triangular matrices.
\end{theorem}

Now, we briefly introduce Picard-Vessiot Theory. First, we say that $\left( \mathcal{K}, \phantom{i}' \ \right)$ - or, simply,
$\mathcal{K}$ - is a \emph{differential field} if $\mathcal{K}$ is a commutative field
of characteristic zero, depending on $x$ and $\phantom{i}'$ is a
derivation on $\mathcal{K}$ (that is, satisfying that $(a+b)'=a'+b'$
and $(a\cdot b)'=a'\cdot b+a \cdot b'$ for all $a,b \in \mathcal{K}$). We
denote by $k$ the \emph{field of constants of $\mathcal{K}$},
defined as $k=\left\{ c \in \mathcal{K} \ | \ c'=0
\right\}$.\\

\noindent We deal with second order linear homogeneous differential
equations, i.e., equations of the form
\begin{equation}
\label{soldeq}
y''+b_1y'+b_0y=0,\quad b_1,b_0\in \mathcal{K},
\end{equation}
and we are concerned with the algebraic structure of their solutions. Moreover, along this work, we refer
to the current differential field as the smallest one containing the field of coefficients of this differential equation.\\

\noindent Let us suppose that $y_1, y_2$ form a basis of solutions for the Equation
\eqref{soldeq}, i.e., $y_1, y_2$ are linearly independent over $k$
and every other solution is a $k$-linear combination of
$y_1$ and $y_2$. Let $\mathcal{L}= \mathcal{K}\langle y_1, y_2 \rangle =\mathcal{K}(y_1, y_2, y_1',
y'_2)$ be the differential extension of $\mathcal{K}$ such that $k$
is the field of constants for both $\mathcal{K}$ and $\mathcal{L}$. In this terms, we say
that $\mathcal{L}$, the smallest differential field containing $\mathcal{K}$ and
$\{y_{1},y_{2}\}$, is the \textit{Picard-Vessiot extension} of $\mathcal{K}$
for the differential equation given in Equation \eqref{soldeq}.\\

\noindent The group of all the differential automorphisms of $\mathcal{L}$ over $\mathcal{K}$
that commute with the derivation $\phantom{i}'$  is called the
\emph{differential Galois group} of $\mathcal{L}$ over $\mathcal{K}$, which is denoted by ${\rm
	DGal}(\mathcal{L}/\mathcal{K})$. This means in particular that for any $\sigma\in
\mathrm{DGal}(\mathcal{L}/\mathcal{K})$, $\sigma(a')=(\sigma(a))'$ for all $a\in \mathcal{L}$
and that $\sigma(a)=a$ for all $a\in \mathcal{K}$. Thus, if $\{y_1,y_2\}$ is
a fundamental system of solutions of Equation ~\eqref{soldeq} and $\sigma \in
\mathrm{DGal}(\mathcal{L}/\mathcal{K})$ then $\{\sigma y_1, \sigma y_2\}$ is also a
fundamental system. This implies the existence of a non-singular
constant matrix
\[
A_\sigma=
\begin{pmatrix}
a & b\\
c & d
\end{pmatrix}
\in \mathrm{GL}_2(\mathbb{C}),
\]
such that
\[
\sigma
\begin{pmatrix}
y_{1}&
y_{2}
\end{pmatrix}
=
\begin{pmatrix}
\sigma (y_{1})&
\sigma (y_{2})
\end{pmatrix}
=\begin{pmatrix} y_{1}& y_{2}
\end{pmatrix}A_\sigma.
\]
This fact can be extended in a natural way to a system
\[
\sigma
\begin{pmatrix}
y_{1}&y_2\\
y'_1&y'_{2}
\end{pmatrix}
=
\begin{pmatrix}
\sigma (y_{1})&\sigma (y_2)\\
\sigma (y'_1)&\sigma (y'_{2})
\end{pmatrix}
=\begin{pmatrix} y_{1}& y_{2}\\y'_1&y'_2
\end{pmatrix}A_\sigma,
\]
which leads to a faithful representation $\mathrm{DGal}(\mathcal{L}/\mathcal{K})\to
\mathrm{GL}_2(\mathbb{C})$ and makes feasible to consider
$\mathrm{DGal}(\mathcal{L}/\mathcal{K})$ as a subgroup of $\mathrm{GL}_2(\mathbb{C})$
depending (up to conjugacy) on the choice of the fundamental system $\{y_1,y_2\}$.\\

\noindent One of the fundamental results of the Picard-Vessiot Theory is the
following theorem (see~\cite{ka}).

\begin{theorem}  The Galois group $\mathrm{DGal}(\mathcal{L}/\mathcal{K})$ is an
	algebraic subgroup of $\mathrm{GL}_2(\mathbb{C})$.
\end{theorem}
We say that the Equation~\eqref{soldeq} is \emph{integrable} if the Picard-Vessiot extension
$\mathcal{L}\supset \mathcal{K}$ is obtained as a tower of differential fields
$\mathcal{K}=\mathcal{L}_0\subset \mathcal{L}_1\subset\cdots\subset \mathcal{L}_m=\mathcal{L}$ such that
$\mathcal{L}_i=\mathcal{L}_{i-1}(\eta)$ for $i=1,\ldots,m$, where either
\begin{itemize}
	\item[$(i)$] $\eta$ is {\emph{algebraic}} over $\mathcal{L}_{i-1}$, that is $\eta$ satisfies a
	polynomial equation with coefficients in $\mathcal{L}_{i-1}$;
	\item[$(ii)$] $\eta$ is {\emph{primitive}} over $\mathcal{L}_{i-1}$, that is $\eta' \in \mathcal{L}_{i-1}$; and
	\item[$(iii)$] $\eta$ is {\emph{exponential}} over $\mathcal{L}_{i-1}$, that is $\eta' /\eta \in \mathcal{L}_{i-1}$.
\end{itemize}

Usually in terms of differential algebra's terminology, we say that
the Equation~\eqref{soldeq} is integrable if the corresponding
Picard-Vessiot extension is \emph{Liouvillian}. Moreover, the
following theorem  holds.

\begin{theorem}[Kolchin]
	\label{LK}
	The Equation~\eqref{soldeq} is integrable if
	and only if $\mathrm{DGal}(\mathcal{L}/\mathcal{K})$ is virtually solvable, that is, its identity component
	$(\mathrm{DGal}(\mathcal{L}/\mathcal{K}))^0$ is solvable.
\end{theorem}

For instance, for the case $b_1=0$ in the Equation~\eqref{soldeq}, i.e.
$y''+b_0y=0$, it is very well known that
$\mathrm{DGal}(\mathcal{L}/\mathcal{K})$ is an algebraic subgroup of ${\rm
	SL}_2(\mathbb{C})$, see \cite{ka,vasi}. In particular, $\mathbb{G}_a$ (the additive group) and $\mathbb{G}_m$ (the multiplicative group) are algebraic subgroups of ${\rm SL}(2,\mathbb{C})$, which are given by 
$$\mathbb{G}_a=\left\{\begin{pmatrix} 1&\mu\\0&1
\end{pmatrix},\quad \mu\in \mathbb{C}\right\},\quad \mathbb{G}_a=\left\{\begin{pmatrix} \lambda&0\\0&\lambda^{-1}
\end{pmatrix},\quad \lambda \in \mathbb{C}^*\right\}.
$$

\bigskip

We recall that $x=x_0\in\mathbb{C}$ is an ordinary point of the Equation \eqref{soldeq} whether $b_0$ and $b_1$ are analytics in $x_0$. On the other hand, we say that $x_0$ is a singular point of the Equation \eqref{soldeq} whether $x_0$ is not an ordinary point for \eqref{soldeq}. Moreover, if $x_0$ is a singular point of the Equation \eqref{soldeq}, we say that it is regular provided that $(x-x_0)^2b_0$ and $(x-x_0)b_1$ are analytic functions of the Equation \eqref{soldeq}. In case that a point is neither ordinary nor singular regular, we say that such point is singular irregular. To study $x=\infty$ as an ordinary point or singular (regular or irregular) point of the Equation \eqref{soldeq}, we make the change of variable $x=z^{-1}$ and then we study the behaviour of $z=0$ in the new equation. This process is a particular case of the \emph{Hamiltonian Algebrization Procedure}. For further details see \cite{acthesis,acbl}.\\

\noindent One important application of differential Galois theory is \emph{Morales Ramis Theory}.
Let us consider a $n$ degrees of freedom
Hamiltonian $H$, given by \begin{equation}\label{ham}H=\frac{||\textbf{p}||^2}{2}+V(\textbf{q}),\quad \textbf{q}=\begin{pmatrix}
q_1\\\vdots\\ q_n\end{pmatrix}, \quad \textbf{p}=\begin{pmatrix}
p_1\\\vdots\\ p_n\end{pmatrix},\quad \textbf{q}\in\mathbb{C}^n,\quad \textbf{p}\in \mathbb{C}^n.\end{equation}
The equations of the flow of the Hamiltonian system, in a
system of canonical coordinates $(\textbf{q},\textbf{p})$, are given by
\begin{equation}\label{hameq}
\dot z =X_H,\quad z= (\textbf{q},\textbf{p}),\quad X_H=J_n\nabla H,\quad J_n=\begin{pmatrix}
\textbf{0}_n&I_n\\
-I_n&\textbf{0}_n
\end{pmatrix},
\end{equation}
where $\textbf{0}_n$ is the square null matrix of size $n^2$, $I_n$ is the identity matrix of size $n^2$, the matrix $J_n$ is known as the \emph{symplectic matrix}, and the vector field $X_H$ is known as the \emph{Hamiltonian vector field}. The system \eqref{hameq} is known as the  \emph{Hamilton equations}, which are conventionaly written as 
$$	\dot q_i = \frac{\partial H}{\partial p_i}, \quad \dot
p_i = - \frac{\partial H}{\partial q_i},\quad i=1,\ldots,n.$$ For instance the Hamiltonian vector field is then written as
$$X_H=\begin{pmatrix}
p_1\\
\vdots\\
p_n\\
-\frac{\partial V(\textbf{q})}{\partial q_1}\\
\vdots\\
-\frac{\partial V(\textbf{q})}{\partial q_n}
\end{pmatrix}.$$ 
Assume that $\gamma=\gamma(t)=(q_1(t),\ldots,q_n(t),p_1(t),\ldots,p_n(t))$ is a particular integral curve (also known as particular solution) of the Hamiltonian system \eqref{hameq}. The Jacobian of the Hamiltonian vector field $X_H$ restricted to $\gamma$, which is exactly the product between the symplectic matrix and the Hessian of the Hamiltonian $H$, is the matrix of coefficients for the \emph{variational equation} of the Hamiltonian system \eqref{hameq}, which is given by
\begin{equation}\label{VE}
\dot \eta=A(t)\eta,\quad \eta=\begin{pmatrix}
\eta_1\\ \vdots\\ \eta_{2n}\end{pmatrix},\quad A(t)=X'_H(\gamma)=J\textrm{Hess}(H)|_\gamma.
\end{equation}

The matrix $A(t)$ can be written explicitly as
$$A(t)\begin{pmatrix}
\textbf{0}_n& I_n\\
-B_n & \textbf{0}_n
\end{pmatrix},\quad B_n=\begin{pmatrix}
\frac{\partial^2V(\textbf{q})}{\partial q_1^2}&\cdots& \frac{\partial^2V(\textbf{q})}{\partial q_n\partial q_1}\\ \vdots\\ \frac{\partial^2V(\textbf{q})}{\partial q_1\partial q_n}&\cdots& \frac{\partial^2V(\textbf{q})}{\partial q_n^2}
\end{pmatrix}.$$
We recall that the \emph{Poisson bracket} between $f(\textbf{q},\textbf{p})$ and $g(\textbf{q},\textbf{p})$ is given by
$$\{f,g\}=\sum_{k=1}^n\left({\partial f\over \partial p_k}{\partial g\over \partial q_k}-{\partial f\over \partial p_k}{\partial g\over \partial q_k}\right).$$ We say that $f$ and $g$ are in \emph{involution} when $\{f,g\}=0$, also we say in this case that $f$ and $g$ commute under the Poisson bracket.\\

\noindent A Hamiltonian $H$ in $\mathbb{C}^{2n}$ is called \emph{integrable in the sense of Liouville} provided that
there exist $n$ independent first integrals of the Hamiltonian system in involution. Following \cite{mo,mora1,mora2,mora3}, we 
say that $H$ in integrable \emph{by terms of rational functions} if we can
find a complete set of integrals within the family of rational functions.
Respectively, we can say that $H$ is integrable \emph{by terms of
	meromorphic functions} if we can find a complete set of integrals within the
family of meromorphic functions.
\emph{Morales-Ramis theory} relates integrability of Hamiltonian systems in the Liouville sense with integrability  Picard-Vessiot theory in terms of differential Galois theory. The following theorem is known as \emph{Morales-Ramis theorem}, see \cite{mo,mora1,mora2,mora3}. 

\begin{theorem}[Morales-Ramis]\label{mora}
	\label{:MR} Let $H$ be a Hamiltonian in $\mathbb{C}^{2n}$, and $\gamma$ be a
	particular solution such that the variational has regular (resp. irregular)
	singularities at the points of $\gamma$ at infinity. If $H$ is
	integrable by terms of meromorphic (resp. rational) functions,
	then the differential Galois group of the variational equation is virtually abelian.
\end{theorem}
This theorem has been applied to prove non-integrability in some physical problems, see \cite{acsiam,aabd,acalde,acbl} and references therein.
We recall that any linear Hamiltonian system is integrable and its corresponding Hamiltonian has the form
\begin{equation}\label{hamquad}
H=\frac{||\textbf{p}||^2}{2}+\textbf{q}^TM\textbf{q},
\end{equation}
where $M$ is a $n^2$ size matrix with entries in $\mathbb{C}$. In particular, we are interested in the case where $M$ is a diagonal matrix, i.e., our main object is a Hamiltonian with uncoupled harmonic oscillators.

\subsection{Representations of quivers}

A \emph{quiver}  $\Gamma$ is a quadruple $\Gamma=(\Gamma_0,\Gamma_1, s, e)$, where $\Gamma_0$ and $\Gamma_1$ are sets, $\Gamma_0$ is called \emph{the set of vertices} and $\Gamma_1$  \emph{the set of arrows} (between vertices) and $s,e: \Gamma_1 \longrightarrow \Gamma_0$ are maps given by $s(\alpha )=i$ and $e(\alpha )=j$ for all arrow $\alpha : i \longrightarrow j$, that is, $\alpha$ is an arrow from the vertex $i$ to the vertex $j$. A \emph{path} in $\Gamma$ is a sequence of arrows $p=\alpha_n \cdots \alpha_1$ such that $e(\alpha_t)=s(\alpha_{t+1})$ for $1 \leq t \leq n-1$ (or  $p_1=(j|\alpha_n
\cdots \alpha_1|i)$ to mean that $p_1$ is a path from the vertex $i$ to the vertex $j$). We denote $l(p)=n$ the \emph{length} of the arrow $p$. A \emph{path of length zero} (or \emph{trivial path}) is a path without arrows and for each vertex $i \in \Gamma_0$, we associate a trivial path denoted by $\epsilon_i$, such that $s(\epsilon_i)=e(\epsilon_i)=i$. \\

Let $k$ be a field and $\Gamma$ a \emph{f\mbox inite quiver}, that is, the sets $\Gamma_0$ and $\Gamma_1$ are f\mbox inite. For $k$ and $\Gamma$, we define an algebra, which we will denote by $k\Gamma$. Let $B$ be the set of all paths in $\Gamma$, and we take $k\Gamma$ the $k-$vector space with basis $B$. Take two paths in $B$, namely  $p_1=(j|\alpha_n
\cdots \alpha_1|i)$ and $p_2=(r|\beta_m \cdots \beta_1|l)$. We define the multiplication of $p_1$ and $p_2$ as follow:

\begin{displaymath}
p_1 \cdot p_2= \left\{ \begin{array}{ll}
(j|\alpha_n \cdots \alpha_1 \beta_m \cdots \beta_1|l), &
\textrm{whether}\,\,i=r\\
0 & \textrm{otherwise}
\end{array} \right\}.
\end{displaymath}              

\noindent The multiplication in $k\Gamma$ is extended linearly. The $k$-algebra $k\Gamma$ is called \emph{the path algebra} of $\Gamma$. We denote by $f.d.(k\Gamma )$ the category of finitely generated modules  over the path algebra $k\Gamma$ and by $J$ the ideal of $k\Gamma$ generated by all arrows of $\Gamma$. An ideal $I$ of $k\Gamma$ is \emph{admissible} if there exists $n>0$ such that $J^n \subset I \subset J^2$. 

It is well-known that many algebras can be expressed as path algebras. More precisely, if $A$ is an indecomposable finite basic algebra over an algebraically closed field $k$, then there is a quiver $\Gamma_A$ and an epimorphism of algebras $\phi:k\Gamma_A \longrightarrow A$ such that $ker\phi$ is an admissible ideal of $k\Gamma$ (Gabriel's Theorem).\\

\noindent Given a field  $k$ and a finite quiver $\Gamma$, we can associate  representations and the corresponding morphisms to obtain the so called  representation category of $\Gamma$, usually denoted by $rep\Gamma$. It is well-known that if $k$ is a field and $\Gamma$ is a finite quiver, then $rep\Gamma$ and $f.d.(k\Gamma)$ are equivalent categories.\\

\noindent Now we describe the representations for the quiver 
$$\begin{array}{lll}
\Gamma^n: &  \xymatrix{
	\bullet \ar@{-}[r] &  \bullet \ar@{-}[r] & \bullet \ar@{-}[r] & \cdots \ar@{-}[r] & \bullet \ar@{-}[r]     & \bullet \ar@{-}[r] & \bullet  } & (n\geq 1),
\end{array}$$ where the edges can be changed by arrows in any direction.
If $p=\alpha_n\cdots \alpha_1\in \Gamma^n$ is a path of length $n\geq 1$, then there exists an indecomposable finite generated $k\Gamma$-module $M[p]$, which can be  described as follows. There is an ordered $k$-basis $\{z_0,z_1,\ldots,z_n\}$ of $M[p]$ such that the action of $k\Gamma$ on $M[p]$ is given by the following representation $$\varphi_p:k\Gamma \to M_{n+1}(k).$$ Let $\text{\bf v}(j)=\mathbf{t}(\alpha_{j+1})$ for $0\leq j\leq n-1$ and $\text{\bf v}(n)=\mathbf{s}(\alpha_n)$. Then for each vertex $i\in \Gamma^n_0$ and for each arrow $\gamma\in \Gamma^n_1$ and for all $0\leq j\leq n$ define
\begin{align*}
\varphi_p(i)(z_j)=\begin{cases}
z_j, &\text{ if $\text{\bf v}(j)=i$}\\
0, &\text{ otherwise}
\end{cases}
&&
\text{ and }
&&
\varphi_p(\gamma)(z_j)=\begin{cases}
z_{j-1}, &\text{ if $\alpha_j=\gamma$}\\
0, &\text{ otherwise}
\end{cases}
\end{align*}

We call $\varphi_p$ the {\it canonical representation} and $\{z_0,z_1,\ldots,z_n\}$ a {\it canonical $k$-basis} for $M[p]$. In particular, in this paper we consider $k=\mathbb{C}$.

\section{Main results}
The results presented in this paper are concerned to the link between differential Galois theory, representation of quivers and seismic models where the square of slowness has constant hessian and there is not interaction between the positions.

According to the Hamiltonian given in \eqref{ham}, the Eikonal equation \eqref{eikonal2} can be seen as a subsystem of the  Hamiltonian in three degrees of freedom 
\begin{equation}\label{eh1}
H=\frac{||\textbf{p}||^2}{2}-\frac{1}{2c^2(\textbf{q})},
\end{equation}
that is, the potential energy for the Hamiltonian $H$ is given by 
$$V(\textbf{q})=-\frac{1}{2c^2(\textbf{q})},\quad V(q_1,q_2,q_3)\in \mathbb{R}(q_1,q_2,q_3)\subseteq \mathbb{C}(q_1,q_2,q_3).$$ In particular, we consider some special cases present in \cite{cerveny2001}, which are summarized in the following result. 

\begin{theorem}\label{theo}
	For a seismic model with constant Hessian of square of slowness without interaction of positions the following statements holds:
	\begin{enumerate}
		\item The differential Galois group of the variational equation of the Hamiltonian system is isomorphic to a triple tensor product with additive and multiplicative groups of complex numbers.
		\item Let $\Gamma^2$ be the quiver with two vertices ($\epsilon_0$ and $\epsilon_1$) and one arrow $\alpha$ from the vertex $\epsilon_1$ to vertex $ \epsilon_0$. If $\varphi_{\alpha}:\mathbb{C}\Gamma \to M_2(\mathbb{C})$ is the representation for the arrow $\alpha$, then $\mathrm{exp}(t\varphi_{\alpha}(\alpha))=\mathbb{G}_a$ and $\mathrm{exp}(t (\epsilon_0-\epsilon_1))=\mathbb{G}_m$.
	\end{enumerate}
\end{theorem}

\textbf{Proof.} We proceed according to each item.
\begin{enumerate}
	
	\item By the Equation \eqref{eh1} and the hypothesis, $Hess(1/c^2(q_1,q_2,q_3))$ is constant. Since $$Hess \left(\frac{1}{2}\left(p_1^2+p_2^2+p_3^2\right)\right)$$ is constant, we get that $Hess(H)$ is also constant. On the other hand, by the Equation \eqref{VE}, we see that the coefficient matrix of the variational equation is constant and is given by $$A=X_H'=J_3Hess(H).$$  For instance, by the Equation \eqref{hameq}, the Hamilton equations form a linear differential system given by $$\dot z=J\nabla H=X_H.$$ Therefore the variational equation and the Hamiltonian system are equivalent. By the Equation \eqref{hamquad} and because there is not interaction between the positions, the Hamiltonian is given by $$H=\frac{1}{2}||\textbf{p}||^2+\textbf{q}^T\textrm{diag}(b_1,b_2,b_3)\textbf{q},$$ i.e., equivalently $$H=\frac{p_1^2+p_2^2+p_3^2}{2}+b_1q_1^2+b_2q_2^2+b_3q_3^2,\quad b_i\in\mathbb{C}.$$ This Hamiltonian can be written as $H=H_1+H_2+H_3$, which is integrable owing to the rational first integrals $H_1$, $H_2$ and $H_3$ are in involution, i.e., $\{H_1,H_2\}=\{H_1,H_3\}=\{H_2,H_3\}=0$, and they are also independent. Since the Hamiltonian system is linear, we obtain general solutions $\gamma=\gamma(t)$ and for instance,  $X'_{H}|_{\gamma}=X_H'$. For instance, the variational equation is equivalent to a linear differential system of three uncoupled second order differential equations with constant coefficients is given by $$\ddot\xi_1=b_1\xi_1,\quad \ddot\xi_2=b_2\xi_2,\quad \ddot\xi_3=b_3\xi_3.$$ These second order differential equations have the same differential field $\mathcal{K}=\mathbb{C}$, which is the field of complex numbers. Furthermore, they have only one singularity namely, $t=\infty$, which is of irregular type due after the change of variable $t=\frac{1}{z}$, $z=0$ is singular irregular point of  $$y''+\frac{2}{z}y'+\frac{b_i}{z^4}y=0.$$ By Morales-Ramis theory, the differential Galois group of the variational equation must be virtually abelian, which in our case corresponds to connected groups computed as follows. To obtain the Picard-Vessiot extensions $\mathcal L$ we can consider two cases for each second order differential equation. The first case corresponds to $b_i=0$, for some $i\in\{1,2,3\}$. Thus, the basis of solutions is given by $\mathcal{B}=\langle 1,t\rangle$ and the Picard-Vessiot extension is given by $\mathcal{L}=\mathbb{C}(t)$. Thus, the differential automorphisms $\sigma:\,\, \mathcal L \rightarrow \mathcal L$ such that $\sigma|_{\mathcal K}=id$ are given by $$\sigma (1)=1,\quad and\quad\sigma\left(\frac{d}{dt} t\right)=\frac{d}{dt} t=\frac{d}{dt}\sigma(t),$$ that is, $\sigma(t)=t+\mu$. This implies that $$\mathrm{DGal}(\mathcal L/\mathbb{C})=\{\sigma_\mu: \sigma_\mu(y)=y+\mu, \,\forall y\in \mathcal{B}, \, \forall k\in\mathbb{C}\}\cong \mathbb{G}_a.$$ 
	The second case corresponds to $b_j\neq 0$, for some $j\in\{1,2,3\}$. Thus, the basis of solutions is given by $\mathcal{B}=\langle \mathrm{exp}(\sqrt{b_j}t),\mathrm{exp}(-\sqrt{b_j}t)\rangle$  and the Picard-Vessiot extension is given by $\mathcal{L}=\mathbb{C}(\mathrm{exp}(\sqrt{b_j}t))$. Thus, the differential automorphisms $\sigma:\,\, \mathcal L \rightarrow \mathcal L$ such that $\sigma|_{\mathcal K}=id$ are given by $$\sigma \left(\frac{d}{dt}\ln (\mathrm{exp}(\sqrt{b_j}t))\right)=\frac{d}{dt}\ln(\mathrm{exp}(\sqrt{b_j}t))=\frac{d}{dt}\sigma (\ln (\mathrm{exp}(\sqrt{b_j}t)))$$ $$and\quad  \sigma(\mathrm{exp}(-\sqrt{b_j}t))=\sigma^{-1}\mathrm{exp}(\sqrt{b_j}t),$$that is, $\sigma(y)=\lambda y$, where $y\in\mathcal{B}$. This implies that $$\mathrm{DGal}(\mathcal L/\mathbb{C})=\{\sigma_\lambda: \sigma_\lambda(y)=\lambda y, \,\forall y\in \mathcal{B}, \, \forall \lambda\in\mathbb{C}^{*}\}\cong \mathbb{G}_m.$$ Since there are three second order differential equations, there are eight possibilities for the differential Galois group of the variational equation, which is isomorphic to  $$\mathbb{G}_i\otimes\mathbb{G}_j\otimes \mathbb{G}_k,\quad i,j,k\in\{a,m\}.$$
	\item Consider the quiver $\Gamma^2$: $\xymatrix{
		1 \ar[r]^{\alpha} &  0}$, if we also consider a $\mathbb{C}$-basis $\{z_0,z_1\}$ of $M[\alpha]$ and let $p=\alpha_1$, with $\alpha_1=\alpha$, then $\varphi_{\alpha}(0)(z_0)=z_0$, $\varphi_{\alpha}(0)(z_1)=0$, $\varphi_{\alpha}(1)(z_0)=0$, $\varphi_{\alpha}(1)(z_1)=z_1$, $\varphi_{\alpha}(\alpha)(z_0)=0$, and  $\varphi_{\alpha}(\alpha)(z_1)=z_0$. Therefore we have that $$\varphi_{\alpha}(\epsilon_0)=\begin{pmatrix}
	1 & 0 \\
	0 & 0
	\end{pmatrix},\quad \varphi_{\alpha}(\epsilon_1)=\begin{pmatrix}
	0 & 0 \\
	0 & 1
	\end{pmatrix},\quad and\quad \varphi_{\alpha}(\alpha)=\begin{pmatrix}
	0 & 1 \\
	0 & 0
	\end{pmatrix}.$$
\end{enumerate}
This finishes the proof of the Theorem \ref{theo}. $\blacksquare$
\begin{remark}
	Theorem \ref{theo} provides an algebraic study of seismic models in seismology, where seismic models with constant velocity field are included in a natural way. Moreover, the square of the slowness for these models with null Hessian have linear heterogeneity in all directions, and the Hamiltonian $H$ can be easily separated in three hamiltonians in one degree of freedom. While for a constant non-null Hessian, the square of the slowness is highly heterogeneous and it can be represented as a quadratic polynomial in three variables without interactions between the positions and the Hamiltonian is separable into three hamiltonians as in the previous case.
\end{remark}

\section*{Final Remarks}
In this paper we studied algebraically some seismic models developed in \cite{cerveny2001}. Using the Hamiltonian approach were considered seismic models with constant Hessian of the square of slowness, without interactions between the positions that are equivalent to hamiltonians of three uncoupled harmonic oscillators. Given that the Hamiltonian system is integrable trough rational first integrals, Morales-Ramis Theorem \ref{mora}, we obtained abelian differential Galois groups and the  representation of the quiver, that allow us to obtain such abelian differential Galois groups, for the variational equations of Hamiltonian systems associated to such seismic models.\\

\noindent We restricted to the case of square slowness without interactions, to motivate the readers through examples about the link between seismology, differential Galois theory and representation theory of quivers. This approach can be extended to more general and complicated examples of seismic models such as anisotropic and inelastic seismic models. It is expected that, seismic models with constant Hessian of square of slowness with interactions between the positions should be studied with this approach in future developments.

\section*{Acknowledgements}
The authors were supported by the COLCIENCIAS-ECOPETROL project (Contract RC. No. 0266-2013) and CODI (Estrategia de Sostenibilidad 2016-2017, Universidad de Antioquia, UdeA). The first author thanks the School of Basic and Biomedical Sciences at the Universidad Sim\'on Bol\'{\i}var for the release time to work in this project. The third author thanks the Department of Basic Science at the Universidad de Medell\'{\i}n for the release time to work in this project. 

\bibliographystyle{plain} 
\bibliography{mybibfile}

\begin{thebibliography}{10}

\bibitem{aabd}
P.~Acosta-Hum\'anez, M.~\'Alvarez-Ram\'{\i}rez, D.~Bl\'azquez-Sanz, and
  J.~Delgado.
\newblock Non-integrability criterium for normal variational equations around
  an integrable subsystem and an example: The \uppercase{W}ilberforce
  spring-pendulum.
\newblock {\em Discrete and Continuous Dynamical Systems - Series A (DCDS-A)},
  33(1):965--986, 2013.

\bibitem{acalde}
P.~Acosta-Hum\'anez, M.~\'Alvarez-Ram\'{\i}rez, and J.~Delgado.
\newblock Non-integrability of some few body problems in two degrees of
  freedom.
\newblock {\em Qualitative Theory of Dynamical Systems}, 8(2):209--239, 2009.

\bibitem{acbl2}
P.~Acosta-Hum\'anez and D.~Bl\'azquez-Sanz.
\newblock Hamiltonian systems and variational equations with polynomial
  coefficients.
\newblock {\em Dynamic systems and applications}, 5(1):6--10, 2008.

\bibitem{acbl}
P.~Acosta-Hum\'anez and D.~Bl\'azquez-Sanz.
\newblock Non-integrability of some hamiltonians with rational potentials.
\newblock {\em Discrete and Continuous Dynamical Systems Series B}, 10(2 \&
  3):265--293, 2008.

\bibitem{acblva}
P.~Acosta-Hum\'anez, D.~Bl\'azquez-Sanz, and C.~Vargas-Contreras.
\newblock On hamiltonian potentials with quartic polynomial normal variational
  equations.
\newblock {\em Nonlinear Studies. The international journal}, 16:299--313,
  2009.

\bibitem{acmowe}
P.~Acosta-Hum\'anez, J.J. Morales-Ruiz, and J.-A. Weil.
\newblock Galoisian approach to integrability of \uppercase{S}chrodinger
  \uppercase{E}quation.
\newblock {\em Report on Mathematical Physics}, 67(3):305--374, 2011.

\bibitem{acpan}
P.~Acosta-Hum\'anez and Ch. Pantazi.
\newblock \uppercase{D}arboux integrals for \uppercase{S}chrodinger planar
  vector fields via \uppercase{D}arboux transformations.
\newblock {\em Symmetry, Integrability and Geometry: Methods and Applications
  (SIGMA)}, 8(043):1--26, 2012.

\bibitem{acsu2}
P.~Acosta-Hum\'anez and E.~Suazo.
\newblock Liouvillian propagators, \uppercase{R}iccati equation and
  differential \uppercase{G}alois theory.
\newblock {\em J. Phys. A: Math. Theor.}, 46(45,455203):1--17, 2013.

\bibitem{acsu}
P.~Acosta-Hum\'anez and E.~Suazo.
\newblock Liouvillian propagators and degenerate parametric amplification with
  time-dependent pump amplitude and phase.
\newblock {\em Analysis, Modelling, Optimization, and Numerical Techniques,
  Springer Proceedings in Mathematics \& Statistics}, 121(1):295--307, 2015.

\bibitem{acthesis}
P.~B. Acosta-Hum\'anez.
\newblock {\em Galoisian Approach to Supersymmetric Quantum Mechanics}.
\newblock Phd Thesis, Universitat Polit\`ecnica de Catalunya, available in
  \href{http://arxiv.org/abs/0906.3532}{ArXiv: 0906.3532}, 2009.

\bibitem{acsiam}
P.~B. Acosta-Hum\'anez.
\newblock Nonautonomous \uppercase{H}amiltonian \uppercase{S}ystems and
  \uppercase{M}orales-\uppercase{R}amis \uppercase{T}heory \uppercase{I}. the
  case $\ddot{x}=f(x,t)$.
\newblock {\em SIAM Journal on Applied Dynamical Systems}, 8(1):279--297, 2009.

\bibitem{acbook}
P.~B. Acosta-Hum\'anez.
\newblock {\em Galoisian Approach to Supersymmetric Quantum Mechanics. The
  integrability analysis of the Schrodinger equation by means of differential
  Galois theory}.
\newblock VDM Verlag, Dr Muller, Berlin, 2010.

\bibitem{akss}
P.~B. Acosta-Hum\'anez, S.I. Kryuchkov, E.~Suazo, and S.K. Suslov.
\newblock Degenerate parametric amplification of squeezed photons: Explicit
  solutions, statistics, means and variances.
\newblock {\em Journal of Nonlinear Optical Physics \& Materials}, 24(2,
  1550021):1--27, 2015.

\bibitem{almp}
P.~B. Acosta-Hum\'anez, J.T. L\'azaro, J.~Morales-Ruiz, and Ch. Pantazi.
\newblock On the integrability of polynomial vector fields in the plane by
  means of \uppercase{P}icard-\uppercase{V}essiot theory.
\newblock {\em Discrete and Continuous Dynamical Systems - Series A (DCDS-A)},
  35(5):1767--1800, 2015.

\bibitem{arnold1989}
V.~Arnold.
\newblock {\em Mathematical Methods of Classical Mechanics}, volume~60 of {\em
  Graduate Texts in Mathematics}.
\newblock Springer Verlag, New York, USA, second edition, 1989.

\bibitem{ASS2006}
I.~Assem, D.~D.~Simson, and A.~Skowro\'nski.
\newblock {\em Elements of the Representation Theory of Associative Algebras}.
\newblock London Mathematical Society Student Texts. Cambridge University
  Press, Cambridge, 2006.

\bibitem{ARS}
M.~Auslander, I.~Reiten, and S.~Smal\o.
\newblock {\em Representation theory of \uppercase{A}rtin Algebras}, volume~36
  of {\em Cambridge Studies in Advanced Mathematics}.
\newblock Cambridge University Press, 1995.

\bibitem{bleistein1986}
N.~Bleistein.
\newblock {\em Mathematical methods for wave phenomena}.
\newblock Academic Press, 1986.

\bibitem{bleistein1984}
N.~Bleistein and R.~Handelsman.
\newblock {\em Asymptotic Expansion of Integrals}.
\newblock Dover, New York, USA, second edition, 1984.

\bibitem{cerveny2001}
V.~Cerveny.
\newblock {\em Seismic Ray Theory}.
\newblock Cambridge University Press, Cambridge, UK, 2001.

\bibitem{Pen1998}
J.~de~la Pe\~na.
\newblock {\em Tame algebras and derived categories}.
\newblock XV Escola de \' Algebra. UFRGS, Brasil, Canela-RS, first edition,
  1998.

\bibitem{Evans2010}
L.~Evans.
\newblock {\em Partial Differential Equations}, volume~19 of {\em Graduate
  Studies in Mathematics}.
\newblock American Mathematical Society, Providence, Rhode Island, USA, second
  edition, 2010.

\bibitem{john1982}
J.~Fritz.
\newblock {\em Partial Differential Equations}, volume~1 of {\em Applied
  Mathematical Sciences}.
\newblock Springer Verlag, New York, USA, fourth edition, 1982.

\bibitem{Ga}
P.~Gabriel.
\newblock Unzerlegbare dartellungen i.
\newblock {\em Manuscripta Math}, 6:71--103, 1972.

\bibitem{Ga1}
P.~Gabriel.
\newblock Auslander-\uppercase{R}eiten sequences and representation-f\mbox
  inite algebras.
\newblock {\em Proc. ICRA II (Ottawa, Canada 1979), Lecture Notes in Math.,
  Springer-Verlag}, 831:1--71, 1980.

\bibitem{Gus}
W.~H. Gustafson.
\newblock The history of algebras and their representations.
\newblock {\em Proc. ICRA III (Puebla, Mexico 1980). Lecture Notes in Math.,
  Springer-Verlag}, 944:1--28, 1982.

\bibitem{ka}
I.~Kaplansky.
\newblock {\em An introduction to differential algebra}.
\newblock Hermann, 1957.

\bibitem{ki}
T.~Kimura.
\newblock On \uppercase{R}iemann's equations which are solvable by quadratures.
\newblock {\em Funkcialaj Ekvacioj}, 12(1):269--281, 1969.

\bibitem{ko}
J.~Kovacic.
\newblock An algorithm for solving second order linear homogeneous differential
  equations.
\newblock {\em J. Symbolic Computation}, 2(1):3--43, 1986.

\bibitem{mag}
A.~Magid.
\newblock {\em Lectures on differential Galois theory}.
\newblock University Lecture Series. American Mathematical Society, Providence,
  RI, 1994.

\bibitem{marram}
J.~Martinet and J.P. Ramis.
\newblock Theorie de galois differentielle et resommation.
\newblock {\em Computer Algebra and Differential Equations}, 193(1):117--214,
  1989.

\bibitem{mo}
J.~Morales-Ruiz.
\newblock {\em Differential Galois Theory and Non-Integrability of Hamiltonian
  Systems}.
\newblock Progress in Mathematics. Birkhauser, Basel, 1999.

\bibitem{mora1}
J.~J. Morales-Ruiz and J.-P. Ramis.
\newblock Galoisian obstructions to integrability of hamiltonian systems
  \uppercase{I}.
\newblock {\em Methods Appl. Anal.}, 8(1):33--96, 2001.

\bibitem{mora2}
J.~J. Morales-Ruiz and J.-P. Ramis.
\newblock Galoisian obstructions to integrability of hamiltonian systems
  \uppercase{II}.
\newblock {\em Methods Appl. Anal.}, 8(1):97--112, 2001.

\bibitem{mora3}
J.~J. Morales-Ruiz and J.-P. Ramis.
\newblock Integrability of dynamical systems through differential
  \uppercase{G}alois theory: a practical guide.
\newblock {\em Differential algebra, complex analysis and orthogonal
  polynomials, Contemp. Math., Amer. Math. Soc.}, 509(1):143--220, 2010.

\bibitem{Rauch2012}
J.~Rauch.
\newblock {\em Hyperbolic Partial Differential in Geometrical Optics}, volume
  133 of {\em Graduate Studies in Mathematics}.
\newblock American Mathematical Society, Providence, Rhode Island, USA, first
  edition, 2012.

\bibitem{Re1}
I.~Reiten.
\newblock An introduction to the representation of \uppercase{A}rtin algebras.
\newblock {\em Bull London Math. Soc.}, 17:209--223, 1985.

\bibitem{Rin1984}
C.~M. Ringel.
\newblock {\em Tame Algebras and Integral Quadratic Forms}, volume 1099 of {\em
  Lecture Notes in Mathematics}.
\newblock Springer-Verlag, 1984.

\bibitem{schleicher2007}
J.~Schleicher, M.~Tygel, and P.~Hubral.
\newblock {\em Seismic True-Amplitude Imaging}, volume~12 of {\em SEG
  Geophysical Developments}.
\newblock Society of Exploration Geophysics, Tulsa, OK, USA, first edition,
  2007.

\bibitem{si2}
M.F. Singer.
\newblock An outline of differential galois theory.
\newblock {\em Computer Algebra and Differential Equations}, 121(1):3--58,
  1989.

\bibitem{vasi}
M.~van~der Put and M.~Singer.
\newblock {\em Galois Theory in Linear Differential Equations}.
\newblock Graduate Text in Mathematics. Springer Verlag, New York, 2003.

\bibitem{Zwors2012}
M.~Zworski.
\newblock {\em Semiclassical Analysis}, volume 138 of {\em Graduate Studies in
  Mathematics}.
\newblock American Mathematical Society, Providence, Rhode Island, USA, first
  edition, 2012.

\end{thebibliography}

\end{document}